\documentclass[manuscript]{aastex61}
\newcommand{\affOSU}{Department of Astronomy, Ohio State University, 140 W. 18th Ave., Columbus, OH 43210, USA}
\newcommand{\affOAUW}{Warsaw University Observatory, Al. Ujazdowskie 4, 00-478 Warszawa, Poland}
\newcommand{\affWARWICK}{Department of Physics, University of Warwick, Coventry CV4 7AL, UK}

\received{XXX}
\revised{XXX}
\accepted{XXX}
\submitjournal{ApJ}

\shorttitle{OGLE-2011-BLG-0173 analysis}
\shortauthors{Poleski et al.}

\begin{document}

\title{An Ice giant exoplanet interpretation of the anomaly\\ in microlensing event  OGLE-2011-BLG-0173}

%% The \author command is the same as before except it now takes an optional
%% argument which is the 16 digit ORCID. The syntax is:
%% \author[xxxx-xxxx-xxxx-xxxx]{Author Name}
%%
%% The new \altaffiliation can be used to indicate some secondary information such as fellowships. 
%% NOTE that if an \altaffiliation command is used it must come BEFORE the \affiliation call,
%% right after the \author command, in order to place the footnotes in the proper location.
%%
%% Use \email to set provide email addresses. Each \email will appear on its own line so you can put multiple email address in one \email call. 
%% A new \correspondingauthor command is available in V6.1 to identify the corresponding author of the manuscript. 

\correspondingauthor{Rados\l{}aw Poleski}
\email{poleski.1@osu.edu}

\author[0000-0002-9245-6368]{Rados\l{}aw Poleski} 
\affiliation{\affOSU}
\affiliation{\affOAUW}
\author[0000-0003-0395-9869]{B.~S.~Gaudi}
\affiliation{\affOSU}
\author{A.~Udalski}
\affiliation{\affOAUW}
\author{M.~K.~Szyma\'nski}
\affiliation{\affOAUW}
\author{I.~Soszy\'nski}
\affiliation{\affOAUW}
\author[0000-0002-2339-5899]{P.~Pietrukowicz}
\affiliation{\affOAUW}
\author[0000-0003-4084-880X]{S.~Koz\l{}owski}
\affiliation{\affOAUW}
\author[0000-0002-2335-1730]{J.~Skowron}
\affiliation{\affOAUW}
\author{\L{}.~Wyrzykowski}
\affiliation{\affOAUW}
\author{K.~Ulaczyk}
\affiliation{\affOAUW}
\affiliation{\affWARWICK}


\begin{abstract}
We analyze the microlensing event OGLE-2011-BLG-0173, which shows a small perturbation 
at the end of the microlensing event caused by the primary lens. 
We consider both  binary lens and binary source models
and we explore their degeneracies, some of which have not previously 
been recognized.   
There are two families of binary lens solutions, one with a mass ratio
$q\approx4\times10^{-4}$ and a separation $s\approx4.6$ and the other with 
$q\approx0.015$ and $s\approx0.22$, i.e, both have companions in the planetary regime.  
We search for solutions by using Bayesian analysis that includes planet frequency 
as a prior and find that the $s\approx4.6$ family is the preferred one with 
$\approx4~\mathrm{M_{Uranus}}$ 
mass planet on an orbit of $\approx10~\mathrm{AU}$. 
The degeneracies arise from a paucity of information on the anomaly, demonstrating
that high-cadence observations are essential for characterizing 
wide-orbit microlensing planets.  Hence, we predict that the planned WFIRST microlensing 
survey will be less prone to these degeneracies than 
the ongoing ground-based surveys.  We discuss the known low-mass, 
wide-orbit companions and we notice that for the largest projected separations 
the mass ratios are either high 
(consistent with brown dwarf companions) or low (consistent with Uranus analogs), 
but intermediate 
mass ratios (Jupiter analogs on wide orbits) have not been detected to date, 
despite the fact that the sensitivity to such planets should be higher than 
that of Uranus analogs. This is therefore tentative evidence of the existence 
of a massive ice giant desert at wide separations.  On the other hand, given 
their low intrinsic detection sensitivity, Uranus analogs may be 
ubiquitous.  
\end{abstract}


% #############################################################################
\section{Introduction} \label{sec:intro}

Our knowledge of the formation of the outer members of planetary systems is 
very limited.  Even in the case of our own solar system, we lack a full 
understanding of how and when Uranus and Neptune
formed and migrated to their current orbits  
\citep[e.g.,][]{pollack96,thommes99,goldreich04,tsiganis05}.  
Among analogs to the known planets in our solar system, ice giant planets are 
particularly hard to detect using conventional techniques \citep[e.g.][]{kane11}, 
due to their faintness in reflected light or thermal emission, their relatively 
low mass, and, most importantly, their large semi-major axes and  longer periods.   
There are two planet detection 
techniques that in principle allow one to detect planets on wide orbits: direct 
imaging and gravitational microlensing.  Directly detecting ice giant analogs in 
mature systems in reflected light is exceptionally challenging because of their 
low planet-to-star flux ratios. For example, the planet-to-star flux ratio is 
$\sim$few$\times 10^{-11}$ in reflected light for a Uranus/Sun analog.     

The gravitational microlensing 
technique is sensitive to the mass of the light-bending body, thus
providing means for finding planets on very wide orbits even if their 
luminosities are low.  The upcoming Wide-Field Infrared Survey Telescope  
(WFIRST) mission \citep{spergel15,penny18_tmp} will give us unprecedented 
opportunity to study the ice giant planets.  The major disadvantage 
of the microlensing - its unpredictability and non-repetitiveness - will be
overcome by WFIRST by collecting high-cadence (15 min.) continuous observations 
over relatively long periods of time (6 seasons of 72 days each). 
In some cases, WFIRST will observe the planetary anomaly 
but will not observe the peak of the host subevent owing to the
long time separation of the two peaks.  Nevertheless, 
in most cases the basic planetary parameters should be constrained 
thanks to the large number of WFIRST epochs or 
the ground-based observations of the host subevent peak.

While the WFIRST survey will definitely change our understanding of outer 
planets in planetary systems, we can and should use existing facilities to study 
the ice giant planets to the extent
possible.  
The studies that are possible with existing facilities can give us first 
estimates of the ice giant planet occurrence rates and 
can help in developing the WFIRST program
by: 1) identifying optimal survey strategies, as well as precursor and
concurrent observations, and 2) identifying and solving difficulties and degeneracies in 
detecting and characterizing planets via microlensing for such systems.  
Among existing facilities, 
the Korean Microlensing Telescope Network \citep[KMTNet;][]{kim16,kim18b} 
is suited best because of te 24-hour coverage allowed by 
the three identical telescopes at different longitudes.

Here we present an analysis of the anomalous microlensing event OGLE-2011-BLG-0173.  
Several degenerate models are found, and our analysis indicates that the preferred 
explanation of the anomaly is a model with  a single source and a lens system being 
a star with a planet on an extremely wide orbit.
The planet-to-host mass ratio is a few  
times $10^{-4}$ and the on-sky-projected separation is almost five times larger
than the angular Einstein ring radius, which is the typical angular distance at 
which planetary perturbations are found.  These 
properties are similar to OGLE-2008-BLG-092LAb \citep{poleski14c}
and make these two planets unique approximate analogs of solar system ice giants.
We analyze the existing data for OGLE-2011-BLG-0173 in detail and uncover a number of 
degeneracies, implying that high-cadence, 
long-term photometric monitoring is crucial to precisely model events due to ice giant 
analogs.

In the next Section, we present available observations of OGLE-2011-BLG-0173.  
Section 3 presents the microlensing model fitting.  In Section 4, 
we estimate the posterior physical properties and compare degenerate models. 
We discuss occurrence rates of microlensing 
ice giant exoplanets in Section 5.  
A summary of our results is presented in Section 6.

% #############################################################################
\section{Observations}

The microlensing event OGLE-2011-BLG-0173 occurred 
at the equatorial coordinates ${\rm R.A.} = 17^{\mathrm h}57^{\mathrm m}15\fs78$, 
${\rm Dec.} = -28\degr14\arcmin01\farcs9$ and was identified by
the Optical Gravitational Lensing Experiment  
(OGLE) Early Warning System \citep{udalski03}.  
The observations were collected by the 1.3-m Warsaw Telescope situated at 
Las Campanas Observatory, Chile.  The telescope is equipped with a mosaic CCD 
camera with a field of view of $1.4~{\rm deg^2}$ \citep{udalski15b}.  The pixel 
scale is $0\farcs26$ and typical seeing is $1\farcs1$.  The majority of the OGLE 
images are collected in the $I$ band and here we use only these data to fit 
the microlensing model.  The additional $V$-band data do not cover the anomaly 
part of the light curve.  During the 2011 bulge season, 
OGLE observed the OGLE-2011-BLG-0173 field with a cadence 
of $20~{\rm min}$, and 2022 $I$-band epochs were collected that year.  
There were 14 and 12 data points collected during the first and second night of 
the anomaly, respectively. 
We also analyzed 909 epochs from the second half of 
the 2010 bulge season to determine a reliable 
baseline brightness.  There were 27 $V$-band epochs in 2011, which we 
supplemented with 35 epochs from 2010 and 2012.  
The photometry of the OGLE images was performed using the difference image 
analysis technique \citep{alard00,wozniak00}.  
The raw uncertainties in difference image photometry are often underestimated, and 
thus we corrected them following \citet{skowron16a}.

The baseline brightness in the standard photometric system is 
$I=15.968~{\rm mag}$ and $(V-I)=2.147~{\rm mag}$.  There is an additional 
star only $0\farcs55$ away that is fainter by $\Delta I=1.1~{\rm mag}$.  

The event OGLE-2011-BLG-0173 was also independently discovered by 
the Microlensing Observations in Astrophysics (MOA) survey \citep{bond01} 
as MOA-2011-BLG-133.  MOA did not collect any useful 
data during the anomaly due to bad weather \citep{suzuki16} and hence, 
we do not analyze MOA data.  We searched for other 
time-series photometry for this event that was taken during the anomaly, 
but none was found.

% #############################################################################
\section{Light-curve Analysis}

\begin{figure}[t!]
\includegraphics[angle=270,width=\textwidth]{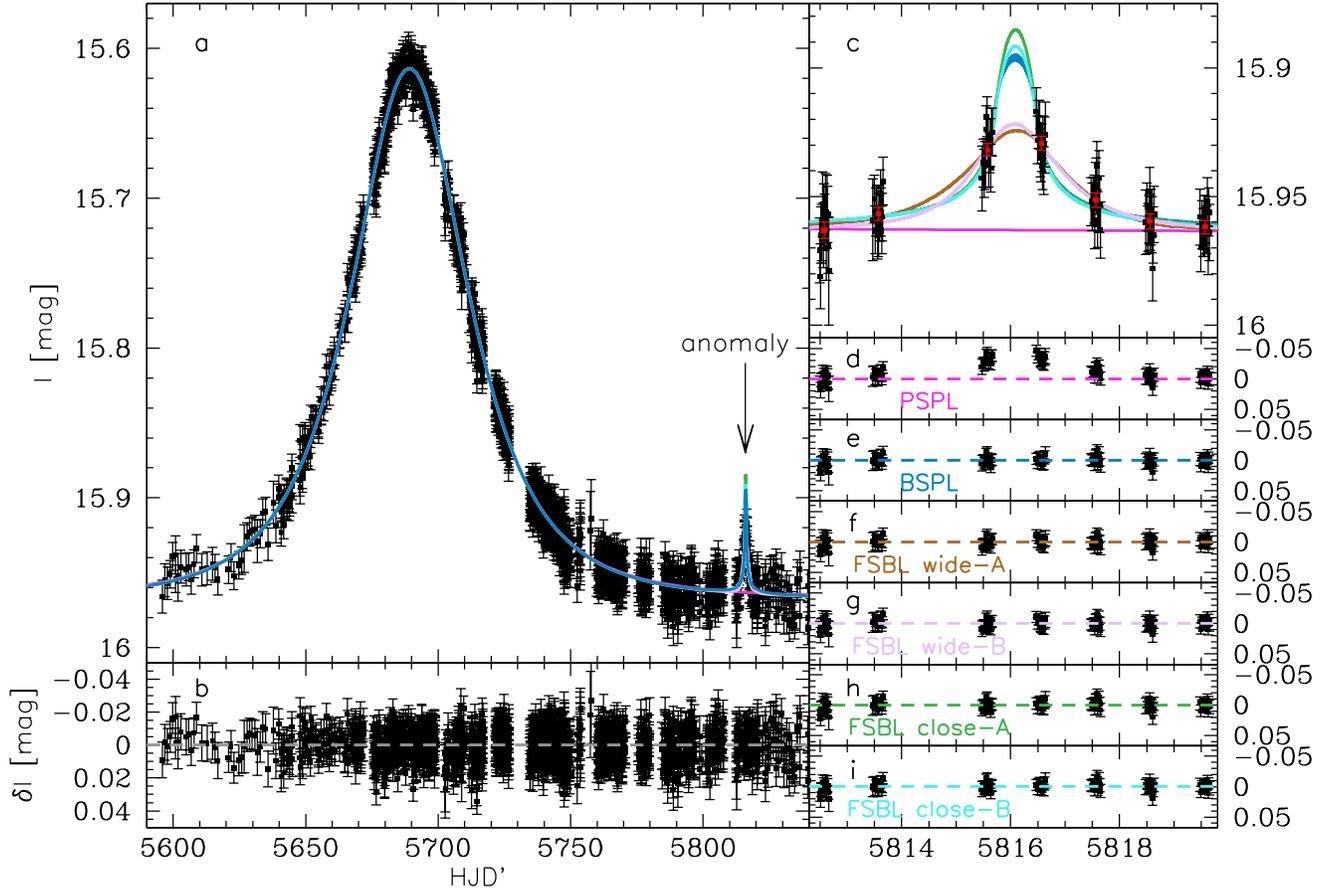}
\caption{Light curve of microlensing event OGLE-2011-BLG-0173. Panel $a$ shows 
the majority of the primary event along with the anomaly. Black points show 
the $I$-band magnitude as a function of ${\rm HJD'} = {\rm HJD}-2450000$ 
in the native cadence of $\sim 20$ minutes. The solid lines are the various 
model fits to the data, described below. 
Panel c shows a zoom-in on the anomaly, where the black points are the native 
sampling and the red points are 
nightly averages of roughly a dozen individual points. 
The colored solid lines indicate the models fit to the anomaly:
PSPL -- magenta, 
binary source -- blue,
binary lens wide-A -- brown,
binary lens wide-B -- lavender,
binary lens close-A -- green, and
binary lens close-B -- cyan.
The residuals from these models are plotted in panels $d$-$i$, respectively. 
Residuals shown in panel $b$ are for the binary lens close model, but 
different models do not differ except for the anomaly.  
The data used to create this figure are available.
\label{fig:lc}}
\end{figure}

The majority of the event light curve appears to be a standard point-source/point-lens (PSPL) 
microlensing light curve \citep{paczynski86} 
with a peak at ${\rm HJD'} \equiv {\rm HJD}-2450000 = 5689$, a timescale of 
$30~{\rm days}$, with the exception of a few-day-long anomaly centered at 
${\rm HJD'} = 5816$ (see Figure~\ref{fig:lc}).  
The PSPL model gives a fit worse by $\Delta\chi^2\approx340$ compared to the more advanced 
models presented below, and 50 consecutive points taken over four nights are 
brighter than the PSPL model.  Hence, the anomaly detection is secure.  
The observed amplitude of the anomaly in the OGLE data is small ($\approx 0.03~{\rm mag}$), 
however, the peak amplitude is not known because it clearly happened when the bulge was not 
observable from Chile.  

The anomalous nature of the event was not recognized while 
the event was ongoing.  This is because the anomaly was brief and 
did not occur when the source was significantly magnified by the primary lens, and thus
the event was close to its baseline brightness and not noticeably magnified 
outside of the two nights when the anomaly occurred. Such short-timescale 
anomalies that occur when the source is not significantly magnified are often 
missed in real time
because of a focus on the anomalies close to the time when events peak.  

The shape of the anomaly appears to be similar to a second point-lens event
with a shorter timescale, either with a point source (PSPL) or 
a finite source (FSPL).
There are two physical situations that can create a microlensing event 
such as OGLE-2011-BLG-0173, which can essentially be described by two well-separated 
brightenings, both of which are consistent with being caused by an isolated lens. 
The first case is if a double source is magnified by a single lens 
\citep{griest92,gaudi98}.  The second case is if a single source is magnified by 
a lens system composed of two bodies.  The possible degeneracy of these two 
scenarios, particularly those for which the two brightenings had very different timescales, 
was recognized and analyzed by \citet{gaudi98}.  However, for OGLE-2011-BLG-0173, each scenario  also 
shows additional degeneracies that have not been identified before.  We discuss the fits to  
the binary source and binary lens models below.  In both cases we tried 
to fit models that included the microlensing parallax effect, but due to 
a large impact parameter ($u_0$) and relatively short Einstein timescale 
($t_{\rm E}$), no meaningful constraints on microlensing parallax 
were found.  Hence, in this case, the microlensing model alone does not lead to 
a direct measurement of the lens mass \citep{gould00b}.

% #############################################################################
\subsection{Binary source model}

The Einstein timescale of the event depends on the lens mass, distances to 
the lens and the source, and the relative lens---source proper motion.  
For a binary source event, all these quantities are the same for both stars, 
hence, the Einstein timescale must be the same for each subevent 
\citep[though see][]{han17e}.  The apparent duration of the second subevent in 
OGLE-2011-BLG-0173 of $\approx 2~{\rm days}$ is much shorter than the duration of the first subevent.  
The subevent duration comparison leads to the conclusion that 
for a double-source model the second source must be 
both significantly more 
blended\footnote{For binary source models, each source is blended by the blending flux and the other source flux.} 
and fainter than the 
first source \citep{gaudi97}.  High blending requires a very close approach of the lens to 
the second source (i.e., $u_{0,2}\ll1$).  The very close approach requires 
including the finite-source effect in the calculations, 
even though the data do not cover the time of minimum separation.  
We evaluated the finite-source magnification based on 
\citet{gould94b} and \citet{yoo04b} prescriptions.  
First, we fitted the model without accounting for the limb darkening.  
The double-source model is characterized by: 
four parameters characterizing the minimum approach to both sources 
(epochs -- $t_{0,1}$ and $t_{0,2}$; separations -- $u_{0,1}$ and $u_{0,2}$)
as well as $t_{\rm E}$, and $\rho_2$ (the radius of the second source 
relative to the characteristic scale of the microlensing event 
$\theta_{\rm E}$).  Additionally, two source fluxes and a blending flux are 
found for each model via linear regression.  
To fit the data, we used the 
Multimodal Ellipsoidal Nested Sampling algorithm, or {\sc MultiNest} 
\citep{feroz08,feroz09}.  

\begin{figure}[t!]
\includegraphics[width=.8\textwidth]{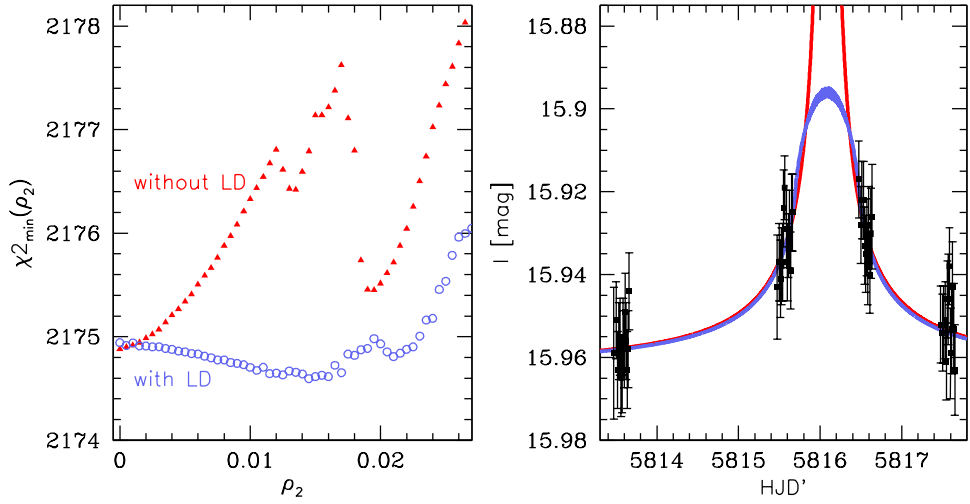}
\caption{ Degeneracy in $\rho_2$ for binary source model. 
{\it Left panel} shows minimum $\chi^2$ calculated in bins of $\rho_2$ that are 0.0005 wide. 
Solid red triangles present models without limb darkening, and 
one can see that at $\rho_{2, {\rm lim}} = 0.0174$ the $\chi^2$ shows a local maximum. 
This produces a  discreet degeneracy in $\rho_2$. 
After including limb darkening (open blue circles) the degeneracy effectively disappears. 
Both curves monotonically rise for  $\rho_2$ values larger than plot limits. 
{\it Right panel} shows the corresponding light curves zoomed in on the anomaly. 
The blue curve is the same as in Figure~\ref{fig:lc}.
\label{fig:minchi2}}
\end{figure}

The first fits that included the finite-source effect and did not include
limb darkening revealed two degenerate binary source models.  The first model 
has $\rho_{2}$ close to zero, and the second one has $\rho_{2}\approx0.019$ 
(see Figure~\ref{fig:minchi2}).  
The $\chi^2$ for both models is similar, but between the two $\rho$ modes the $\chi^2$ values 
are larger by up to $3.1$ and the highest values are found at $\rho_{2, {\rm lim}} = 0.0174$.
The magnification of the second source rises and falls more steeply 
for models without limb darkening than for models with limb darkening. 
However, on neither of the two anomaly nights does the light curve show a significant slope.  
The source diameter crossing time is 
$2\rho_{2}t_{\rm E} = 1.0~{\rm day}$ for $\rho_{2, {\rm lim}}$.
Hence, the models with $\rho_{2}\approx\rho_{2, {\rm lim}}$ show slightly higher 
$\chi^2$ and produce discreet degeneracy. 
Models with $\rho_2 > \rho_{2, {\rm lim}}$ predict 
an unreasonably small mean value and range of
the lens---source relative proper motion: 
$\mu = \theta_{\star,2}/(\rho_2t_{\rm E}) = 0.108 \pm 0.016~{\rm mas~yr^{-1}}$ 
(for determination of angular source radius $\theta_{\star,2}$ see below).  
The two models effectively merge into a single model after the limb-darkening effect is included 
because the rising and falling wings of the anomaly are less steep for 
$\rho_2 > 0$ when finite-source effects are included.  
We assumed a linear limb-darkening coefficient 
of $u = 0.33$.  To find this value, we used \citet{claret11} 
tables and assumed $I$-band absolute brightness of 
$M_{I,2} = 8.3~{\rm mag}$, 
which results from the preliminary fitted 
model and analysis of the color-magnitude diagram (CMD; see details below).  
Table~\ref{tab:bs} gives results of the binary source fitting. 
We separately fitted the model with $\rho_2=0$ and found that 
it is worse by 
$\Delta\chi^2 = 0.31$ compared to the model presented in Table~\ref{tab:bs}.

The amount of blending light we observe, corresponding to $I=17.839~{\rm mag}$, is too large to be fully produced by the lens. 
The blending light could be attributed to a wide-orbit companion to the source or 
to the lens, a star that is not connected to either of them, or any combination 
of those possibilities. 

\begin{deluxetable}{l|r}
\tablecaption{Binary source model parameters\label{tab:bs}}
\tablehead{
\colhead{Parameter} & \colhead{value} 
}
\colnumbers
\startdata
$t_{0,1}$  & $ 5689.189 \pm 0.044 $  \\
$u_{0,1}$  & $ 0.852 \pm 0.039 $  \\
$t_{\rm E}$ (d)  & $ 29.18 \pm 0.86 $  \\
$t_{0,2}$  & $ 5816.100 \pm 0.072 $  \\
$u_{0,2}$  & $ 0.0079 _{- 0.0047 } ^{+ 0.0063 } $  \\
$\rho_2$  & $ 0.0209 _{- 0.0123 } ^{+ 0.0097 } $  \\
$F_{s,1}/F_{\rm base}$\tablenotemark{a}  & $ 0.821 _{- 0.065 } ^{+ 0.079 } $  \\
%$F_{s,2}/F_{\rm base}$\tablenotemark{a}  & $ 0.000536 _{- 0.000060 } ^{+ 0.000075 } $  \\
$F_{s,2}/F_{\rm base}$\tablenotemark{a}  & $ 5.36_{-0.60}^{+0.75}\times10^{-4} $ \\
\hline 
$\chi^2/d.o.f.$  & $2174.60/2922$   \\
\enddata
\tablenotetext{a}{$F_{s,1}$ and $F_{s,2}$ denote fluxes of the two sources, and 
$F_{\rm base}$ denotes the baseline flux (all in the $I$ band).}
%\tablecomments{Note}
\end{deluxetable}

In order to verify  the double-source model we checked for the
chromatic effect predicted by \citet{gaudi98}, and we discuss this possibility below. 

The posterior proper motion distribution without Galactic priors 
is very wide: $\mu=0.53_{-0.39}^{+13.1}~{\rm mas~yr^{-1}}$, hence, 
even if we wait many years and use high angular resolution observations, we may 
not be able to resolve the lens and source and thus infer the proper motion.  
Resolving the two 
source components is currently beyond the capabilities of existing telescopes.  
Therefore, we conclude that continuous, higher-cadence, and higher-precision 
observations of such anomalies are crucial for proving the double-source 
model.  Namely, more data would have provided a higher $\chi^2$ difference 
between the different models, helped to verify whether both subevents have the same source 
color, and better constrained the relative proper motion 
for future verification via high-resolution imaging \citep{gaudi98}.

% #############################################################################
\subsection{Binary lens model}

The main parameters describing the binary lens models are $q$ (the mass ratio of 
the lens components) and $s$ (the projected separation relative to  
$\theta_{\rm E}$).  Binary lenses produce a set of closed curves called caustics, 
on which a point source would have infinite magnification.  Depending on $s$ and $q$, 
the number of caustics can be one, two, or three \citep{schneider86}.  
We can exclude topologies 
with a single caustic, as they cannot produce PSPL light curves with 
a subevent as observed here.  The small ratio of timescales 
for both subevents in OGLE-2011-BLG-0173 strongly suggests a small 
mass ratio: 
$q\approx\left(t_{{\rm E},2}/t_{{\rm E}, 1}\right)^2\approx\left(1~{\rm d}/30~{\rm d}\right)^2\approx0.0011$.  
The long time difference between subevents ($\Delta t$) suggests a large value of 
projected separation: $s \approx \Delta t/t_{\rm E} = 4.3$. 
For the binary lens model, we assumed a limb-darkening coefficient of 
$u=0.64$ based on the preliminary fitted model and \citet{claret11}. 

\begin{figure}[t!]
%\plotone{traj_ob110173_v2.eps}
\plotone{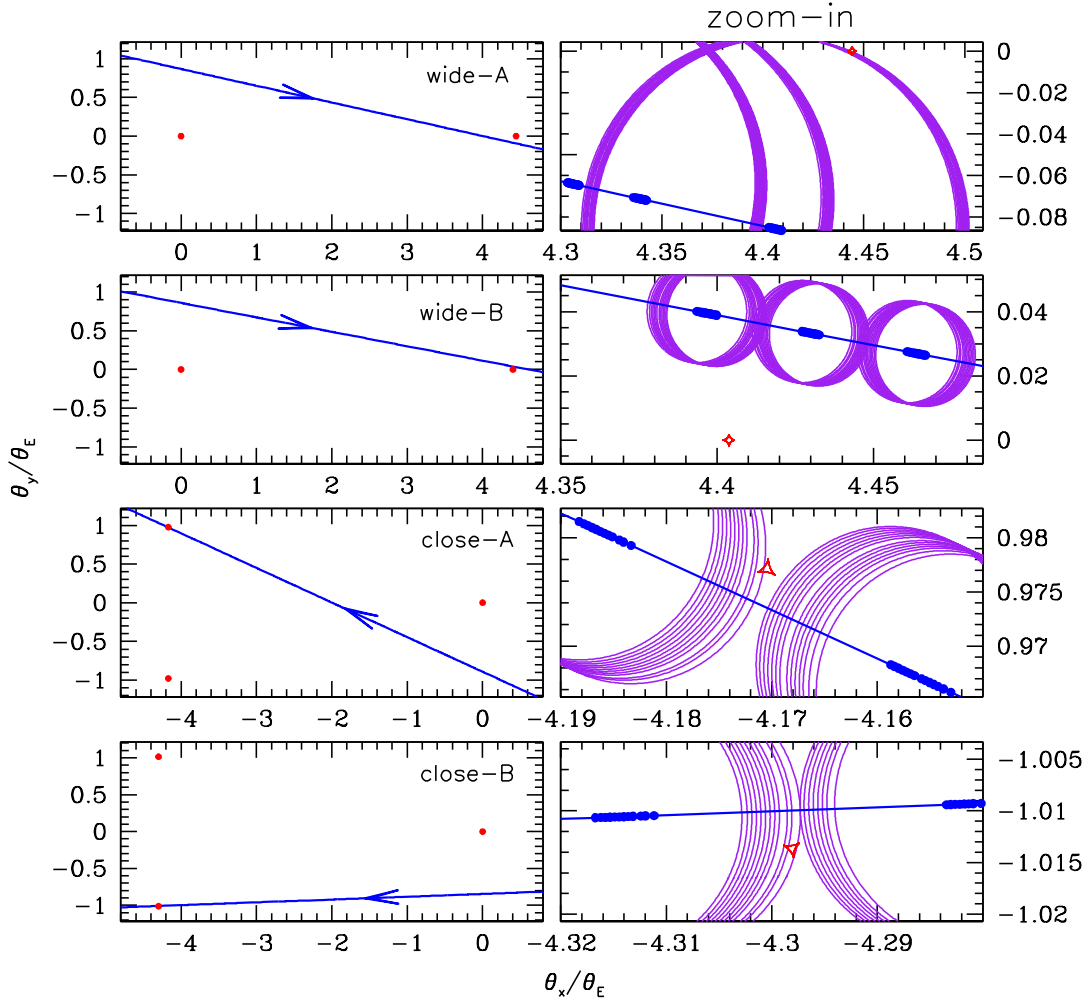}
\caption{Source trajectories and caustics of OGLE-2011-BLG-0173 for different 
binary lens models (one per row).  Each row shows all caustics (large red dots) 
in the left column and zoom-in on the trajectory close to the caustic responsible for the
short-timescale anomaly in the right column.  
The arrows mark the direction of the source motion. In the right column, the large 
blue dots mark the positions of the source during two or three nights close to 
the peak of anomaly, and the purple rings mark the source perimeter for each epoch.
\label{fig:traj}}
\end{figure}

Despite the above arguments, we ran an intensive {\sc MultiNest} simulation to ensure that 
all the degenerate solutions were found.  To our surprise, two families of degenerate models were 
identified -- see Figure~\ref{fig:traj}.  
First, the planetary model that we expected: $s\approx4.6$ 
(wide solution) and 
$q\approx0.0005$ that produces two caustics, with the source approaching 
the planetary caustic, which causes the anomaly.  The source can 
pass both caustics on the opposite or same sides \citep{gaudi97}; 
we denote these models wide-A and wide-B, respectively.
The corresponding models differ significantly 
in only one parameter: the angle 
between source trajectory and binary lens axis ($\alpha$).   
Both models show continuous degeneracy in $\rho$.  
For wide-A the best $\chi^2$ solution has $\rho\approx0.09$, while 
for wide-B $\rho\approx0.016$. 
Second, we found a solution with $s\approx0.22$ 
(close solution) and $q\approx0.014$ that produces 
a central caustic and two planetary caustics, where a close approach to one of 
the two planetary caustics 
creates the anomaly.  This family also further divides into two separate modes 
with the source crossing the binary axis between planetary and central caustics (close-A), 
or beyond the central caustic (close-B). 
The distance between planetary and central caustics projected on 
a binary axis is $\left|s-s^{-1}\right|$ \citep{han06}
and the better approximation is $s(1+q)^2-(s(1+q)^2)^{-1}$ \citep{bozza00}. 
Hence, 
$s\approx4.6$ and $s\approx0.22$ give 
similar values of the caustics distance.  The degeneracy between these two 
families of models was not known before: \citet{gaudi97} explored anomalies in 
the light curves of planetary microlensing events and found that 
the shapes of 
anomalies are different for $s>1$ and $s<1$ at fixed $q$. 
This degeneracy is similar in spirit, although more general, to
the degeneracy between a close binary lens with orbital motion and 
a circumbinary planet \citep{bennett99,albrow00,jung13}. The new 
degeneracy is present in OGLE-2011-BLG-0173 because of poor sampling of 
the anomaly.  
The anomalies in the close solution have smaller amplitude for given $q$, 
hence, significantly larger $q$ is needed for the close model to produce 
the anomaly light curve similar to that of the wide model.
In a typical situation for small values of $q$ and $s<1$ the anomaly
is observed when the source passes between the caustics through 
a demagnification region \citep[e.g., ][]{sumi10}, 
which is absent in the $s>1$ case.  

The parameters of the fitted models are presented in Table~\ref{tab:bl}.  
The last two rows of the table present $\chi^2$ per degree of freedom 
and the Bayesian evidence relative to the close-A model as returned by MultiNest. 

\begin{deluxetable}{l|r|r|r|r}
\tablecaption{Binary lens models parameters\label{tab:bl}}
\tablehead{
\colhead{Parameter} & \colhead{wide-A} & \colhead{wide-B} & \colhead{close-A} & \colhead{close-B}  
}
\colnumbers
\startdata
$t_0$   & $ 5689.190 \pm 0.043 $  & $ 5689.190 \pm 0.043 $  & $ 5689.211 \pm 0.045 $  & $ 5689.139 \pm 0.045 $  \\
$u_0$  & $ 0.850 \pm 0.036 $  & $ 0.851 \pm 0.037 $  & $ 0.817 \pm 0.035 $  & $ 0.844 \pm 0.038 $  \\
$t_{\rm E}$ (d)   & $ 29.20 \pm 0.80 $  & $ 29.19 \pm 0.82 $  & $ 30.15 \pm 0.83 $  & $ 29.37 \pm 0.86 $  \\
$\rho$  & $ 0.064 \pm 0.039 $  & $ 0.070 \pm 0.042 $  & $ 0.0199 _{- 0.0116 } ^{+ 0.0089 } $  & $ 0.0205 _{- 0.0108 } ^{+ 0.0084 } $  \\
$\alpha$ (deg)   & $ 168.05 _{- 0.41 } ^{+ 0.35 } $  & $ 169.86 _{- 0.40 } ^{+ 0.41 } $  & $ 335.10 _{- 1.55 } ^{+ 1.28 } $  & $ 3.21 _{- 1.35 } ^{+ 1.66 } $  \\
$s$  & $ 4.66 \pm 0.12 $  & $ 4.63 \pm 0.13 $  & $ 0.2220 \pm 0.0056 $  & $ 0.2167 _{- 0.0061 } ^{+ 0.0057 } $  \\
$q$   & $ 0.00046 \pm 0.00014 $  & $ 0.00042 \pm 0.00015 $  & $ 0.0144 _{- 0.0025 } ^{+ 0.0034 } $  & $ 0.0151 _{- 0.0028 } ^{+ 0.0040 } $  \\
$F_s/F_{\rm base}$\tablenotemark{a}  & $ 0.818 _{- 0.061 } ^{+ 0.073 } $  & $ 0.818 _{- 0.064 } ^{+ 0.073 } $  & $ 0.755 _{- 0.055 } ^{+ 0.065 } $  & $ 0.805 _{- 0.062 } ^{+ 0.078 } $  \\
\hline 
$\chi^2/d.o.f.$  & $2172.30/2922$ & $2173.54/2922$ & $2167.03/2922$ & $2173.15/2922$   \\
$Z/Z_{\rm close-A}$\tablenotemark{b} & $0.0321$ & $0.0169$ & $1.0000$ & $0.0422$ \\
\enddata
\tablecomments{The parameter conventions follow \citet{skowron11} except 
$t_0$ and $u_0$ which are measured relative to the higher-mass lens component.}
\tablenotetext{a}{Source flux to baseline flux in $I$ band.}
\tablenotetext{b}{Bayesian evidence relative to the close-A solution.}
%\tablecomments{Note}
\end{deluxetable}

\begin{figure}[t!]
%\plotone{max_mag_ob110173_v3.eps}
\plotone{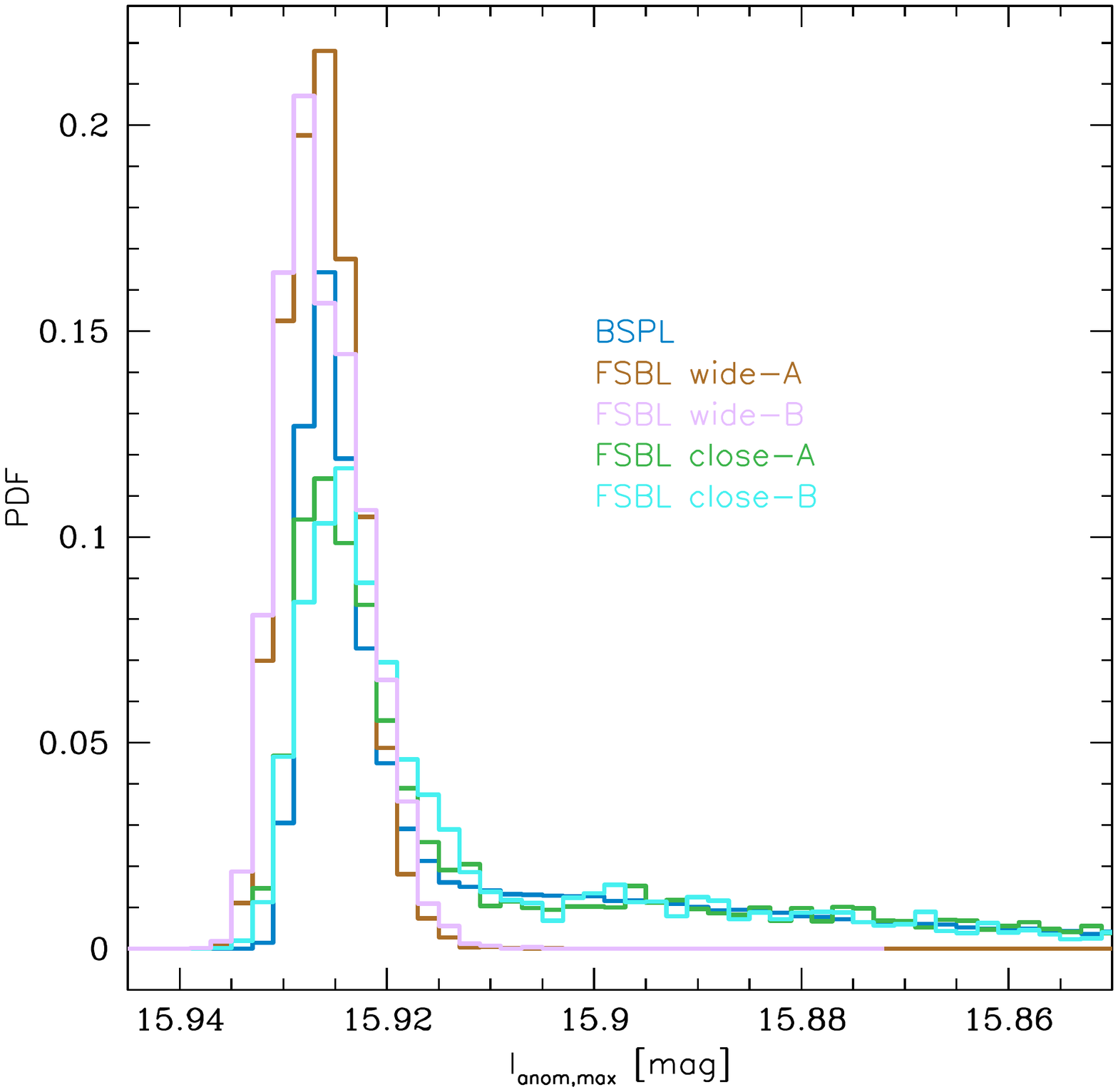}
\caption{Probability distribution of peak brightness during the anomaly 
for five different models. 
\label{fig:mag}}
\end{figure}

Models presented above predict different brightnesses at the peak of 
the anomaly.  Distributions of $I$-band peak brightness are plotted in 
Figure~\ref{fig:mag}.  
The range of peak brightness is wide for binary source and close binary 
lens models. The wide binary lens models predict a narrow range of 
peak brightness. Figure~\ref{fig:mag} shows that a single brightness 
measurement at the peak of the anomaly would not be enough to fully 
resolve the degeneracy. Once more we see that continuous coverage of 
planetary anomaly would help resolve the degenerate models. 

% #############################################################################
\subsection{Source properties}

The physical scale of the microlensing event is set by the Einstein ring 
radius $\theta_{\rm E} =  \theta_{\star}/\rho$. The value of the angular 
source radius $\theta_{\star}$ is estimated here twice. First, for the source 
in the binary lens model ($\theta_{\star,1}$) and 
second, for the second source ($\theta_{\star,2}$) in the binary source model. 

\begin{figure}[t!]
%\plotone{ob110173_cmd_v2.eps}
\plotone{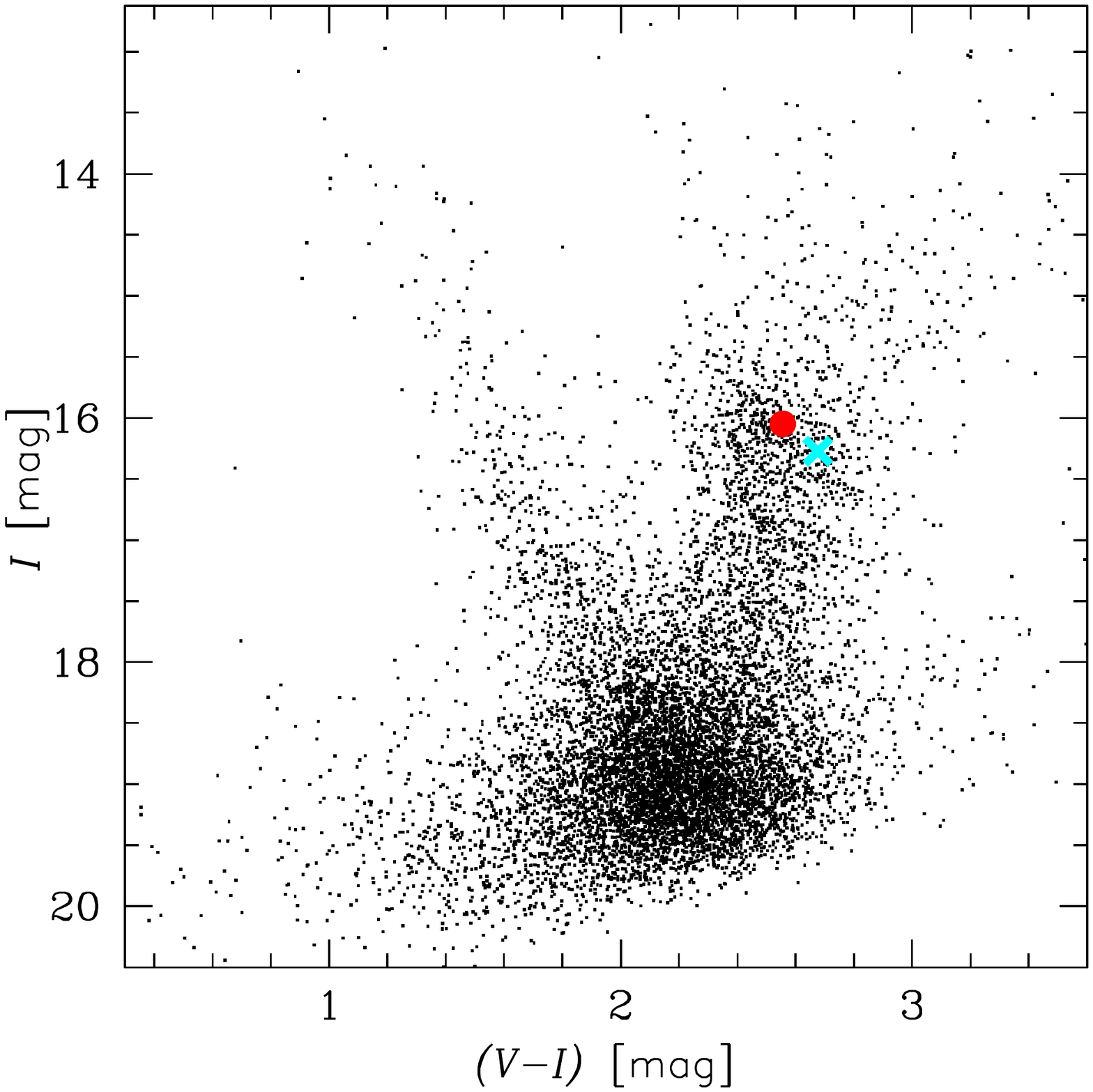}
\caption{Color-magnitude diagram for stars within $2\arcmin$ from the event.  
Red circle marks the red clump centroid.  The cyan cross marks the source position. 
\label{fig:cmd}}
\end{figure}

To derive $\theta_{\star}$ for the binary lens model, we follow the method 
presented by \citet{yoo04b}. Figure~\ref{fig:cmd} presents the CMD for the stars 
close to the event. 
The measured properties of the red clump (RC) centroid are:
$(V-I)_{\rm RC} = 2.556\pm0.006~{\rm mag}$ and 
$I_{\rm RC} = 16.048\pm0.016~{\rm mag}$.
The extinction-free parameters of the RC centroid are:
$(V-I)_{{\rm RC}, 0} = 1.06~{\rm mag}$ \citep{bensby11} and 
$I_{{\rm RC}, 0} = 14.375~{\rm mag}$ \citep[Table 2 of][]{nataf13b}. 
By comparing the above values, we find the reddening and the extinction of 
$E(V-I) = 1.496~{\rm mag}$ and $A_I = 1.673~{\rm mag}$.  
The source unmagnified brightness ($I_s=16.272~{\rm mag}$ and 
$V_s=18.949~{\rm mag}$) is corrected for extinction: 
$I_{s,0} = 14.599~{\rm mag}$ and $V_{s,0} = 15.780~{\rm mag}$. 
The extinction-free magnitudes correspond to the near-infrared brightness of 
$K_{s,0} = 13.063~{\rm mag}$ which we calculated using the intrinsic colors of 
giant stars \citep{bessell88}.
After applying the \citet{kervella04b} relation between $(V-K)$ color 
and the surface brightness, 
we obtain $\theta_{\star,1} = 7.50_{-0.11}^{+0.23}~{\rm \mu as}$.  
For consistency of the results, the above procedure, which effectively 
translates ($I_s$, $V_s$) to $\theta_{\star,1}$, was repeated for every 
model in the posterior distributions.  Based on that, we derive 
the following values of $\theta_{\rm E}$:
$0.36_{-0.25}^{+3.03}~{\rm mas}$ for wide-A,
$0.33_{-0.24}^{+2.43}~{\rm mas}$ for wide-B,
$0.79_{-0.46}^{+4.36}~{\rm mas}$ for close-A, and 
$0.76_{-0.43}^{+4.71}~{\rm mas}$ for close-B model.

For the binary source model we cannot use the \citet{yoo04b} method, because 
only single-band photometry is available for the anomaly. Instead, we derive 
the angular source radius by combining the measured source flux and the stellar 
isochrones.  The source distance modulus is ${\rm DM}_s = 14.481~{\rm mag}$ 
based on the mean bulge distance as a function of Galactic longitude 
\citep{nataf13b}.  Thus, the first-source absolute brightness is 
$M_{I,s} = 0.118~{\rm mag}$.  The flux ratio of the source components corresponds 
to $8.18~{\rm mag}$, hence, the absolute brightness of the second source is
$M_{I,s2} = 8.29~{\rm mag}$.  We extracted the radii for dwarfs 
(age of $10~{\rm Gyr}$, ${\rm [Fe/H]}=-1$, and $Y=0.27$) from \citet{dotter08} 
isochrones.  The radius, combined with the source distance, leads to 
$\theta_{\star,2} = 0.2282\pm0.0096~{\rm \mu as}$ and 
$\theta_{\rm E} = 0.042_{-0.031}^{+1.041}~{\rm mas}$.

% #############################################################################
\section{Interpretation}
\label{sec:inter}

The analysis of the microlensing light curve leads us to five different 
scenarios that could produce the observed signal: single lens with binary 
source or four different scenarios of binary lens with a single source.  
Two binary lens 
scenarios have $(s, q)$ of $(4.6, 0.0005)$ and the other two of $(0.22, 0.014)$.  
We compare these different scenarios by combining Bayesian 
evidence extracted from {\sc MultiNest} 
runs with priors from the Galactic model 
and statistical properties of microlensing planetary systems.  

In order to constrain the prior probabilities, we used the modified version of 
the Galactic model by \citet[][see there for details]{clanton14a}.  The lenses are 
main-sequence stars drawn from density profiles of a double-exponential disk 
and a boxy Gaussian bulge.  The parameters of the main-sequence mass function are 
taken from model 1 of \citet{sumi11}.  The source distance is drawn from 
the boxy Gaussian bulge distribution \citep[model G2 in][]{dwek95}.  
The relative proper motion cutoff was increased to $80~{\rm mas~yr^{-1}}$ in 
each coordinate.  To set the priors, we use only the marginalized event rate
${d^2}\Gamma/{d}t_{\rm E}{d}\theta_{\rm E}$.  

We combined the estimates of $\theta_{\star,1}$ and $\theta_{\star,2}$, 
the measured parameters, and $\Gamma$ values to derive posterior 
statistics of physical event parameters.  These statistics are presented in 
Table~\ref{tab:post} for binary source and all binary lens models.  
The Einstein ring radius in linear units ($r_{\rm E}$) 
values for binary lens models lead to projected separations 
on the order of $8~{\rm AU}$ and $0.5~{\rm AU}$ for wide and close models, 
respectively.  The ice line is at a distance of 
$\sim 2.7(M_l/M_{\odot})~{\rm AU} = 1.1~{\rm AU}$ \citep{kennedy08}, i.e., a few times 
closer to the star than the planet in the wide solution. 

\begin{deluxetable}{l|r|r|r|r|r}
\tablecaption{Posterior physical parameters statistics\label{tab:post}} 
\tablehead{
\colhead{Parameter} & \colhead{BSPL} & \colhead{FSBL wide-A} & \colhead{FSBL wide-B} & \colhead{FSBL close-A} & \colhead{FSBL close-B}
}
\colnumbers
\startdata
$\Gamma/\Gamma_{\rm close-A}$ & 
\nodata & $0.397$ & $0.382$  & $1.000$ &  $1.093$ \\
$\mu~{\rm (mas~yr^{-1})}$ & 
$4.2_{-1.8}^{+2.7}$ & 	$4.3_{-1.8}^{+2.6}$ &	$4.7_{-1.9}^{+2.6}$ &	$4.6_{-1.2}^{+2.7}$ &	$4.7_{-1.2}^{+2.5}$ \\ 
$D_l~{\rm (kpc)}$ & 
$6.2_{-1.7}^{+1.3}$ & 	$6.1_{-1.7}^{+1.3}$ &	$6.0_{-1.7}^{+1.3}$ &	$5.9_{-1.6}^{+1.2}$ &	$5.9_{-1.6}^{+1.2}$\\
$M_l~{\rm (M_{\odot})}$ & 
$0.40_{-0.21}^{+0.30}$ & 	$0.41_{-0.21}^{+0.30}$ & 	$0.42_{-0.22}^{+0.29}$ &	$0.43_{-0.21}^{+0.28}$ &	$0.43_{-0.21}^{+0.28}$\\
$M_c~{\rm (M_{Jup})}$ & 
\nodata & 	$0.19_{-0.11}^{+0.18}$ & 	$0.18_{-0.10}^{+0.17}$ &	$6.5_{-3.2}^{+4.8}$ &	$6.8_{-3.4}^{+5.1}$ \\
$r_{\rm E}~{\rm (AU)}$ & 
$1.91_{-0.65}^{+0.80}$ & 	$1.96_{-0.65}^{+0.80}$ & 	$2.04_{-0.68}^{+0.78}$ &	$2.12_{-0.50}^{+0.68}$ &	$2.12_{-0.50}^{+0.67}$ \\
$\hat{r}_{\rm E}~{\rm (AU)}$ & 
$2.7_{-1.2}^{+1.6}$ & 	$2.9_{-1.2}^{+1.6}$ &	$3.0_{-1.2}^{+1.6}$ &	$3.09_{-0.79}^{+1.57}$ &	$3.06_{-0.77}^{+1.45}$ \\
\enddata
\tablecomments{The parameters are:
$\Gamma/\Gamma_{\rm close-A}$ -- event rate relative to FSBL close-A model, 
$\mu$ -- relative lens-source proper motion, 
$D_l$ -- lens distance, 
$M_l$ -- lens mass, 
$M_c$ -- mass of lens companion (if present), 
$r_{\rm E}$ -- Einstein ring radius (in lens plane), and 
$\hat{r}_{\rm E}$ -- Einstein ring radius projected on the source plane.}
\end{deluxetable}

The parameters $t_{\rm E}$ and $\theta_{\rm E}$ were estimated 
for each model in the posterior distributions, thus allowing integration
of $\Gamma$ over posterior distributions via importance sampling.  
In Table~\ref{tab:post} we provide event rates relative to the close-A model.

The prior on binary lens models additionally depends on the planet properties.  
The statistical properties of microlensing planets were derived 
by \citet{suzuki16} using MOA survey data and combined with earlier results 
by \citet{gould10} and \citet{cassan12}.  
The planet mass ratio and separation function is: 
\begin{equation}
\frac{d^2N_{\rm pl}}{d\log q~d\log s} = A\left(\frac{q}{1.7\times10^{-4}}\right)^n s^m
\end{equation}
where $A=0.61^{+0.21}_{-0.16}$, 
$m=0.49^{+0.47}_{-0.49}$, and 
$n=-0.93\pm0.13$ for $q>1.7\times10^{-4}$ and $n=0.6^{+0.5}_{-0.4}$ otherwise.
The sensitivity of the \citet{suzuki16} study to planets similar to our wide 
model is low, but this study gives the best information we can currently use as 
a prior.  The uncertainty of the coefficient $A$ is large, but it is not important 
when different binary lens models are compared.  Integrating the above 
function results in:
\begin{itemize}
 \item $N_{\rm pl, wide-A}/N_{\rm pl, close-A} = 178$
 \item $N_{\rm pl, wide-B}/N_{\rm pl, close-A} = 212$
 \item $N_{\rm pl, close-B}/N_{\rm pl, close-A} = 0.952$
\end{itemize}

We multiply the microlensing event rate (Table~\ref{tab:post}),
the planet mass ratio and separation functions (above), and 
the Bayesian evidence (Table~\ref{tab:bl}) to find the posterior odds:
\begin{itemize}
 \item wide-A to close-A: $2.27$,
 \item wide-B to close-A: $1.37$,
 \item close-B to close-A: $0.044$.
\end{itemize}
Hence, the wide-A model is preferred.  The posterior odds for the wide model family 
relative to the close model family are $3.5$. 
The preference for wide models over the close models 
comes mostly from the \citet{suzuki16} prior.  Simply
put: we know from previous microlensing surveys that, for planets beyond 
the snow-line (i.e., $s\gtrsim0.3$),  
low-mass planets are much more abundant than high-mass planets, 
and we assume that this holds also for wide-orbit planets. 
Most of the preference in the \citet{suzuki16} prior for wide models 
comes from the mass ratio, not the separation itself. 
The notion that the planets with the smallest mass ratios are more common 
than the higher mass ratio ones was confirmed by a joint analysis of 
microlensing, as well as radial velocity and direct imaging data -- see 
\citet{clanton16}.

It is much more difficult to compare the binary source model versus the binary lens model priors.  
Both models have three of the same parameters that constrain the main subevent 
($t_0$, $u_0$, and $t_{\rm E}$) but differ in other ones: 
four for the binary lens ($\rho$, $\alpha$, $s$, and $q$) vs. three for 
the binary source ($t_{0,2}$, $u_{0,2}$, and $\rho_2$).  We compare 
binary lens and binary source models below.

Binary lens models fit the observed light curve better than the binary source 
model with a $\chi^2$ difference of $2.30$ and $7.57$ relative to the close and 
wide family of models, respectively.  According to \citet{suzuki16}, 
$d^2N_{\rm pl}/(\log q~d\log s)\approx0.5$ for the wide model.  
Also, stellar companions are present around half of the stars 
\citep{raghavan10} but this includes all separations and mass ratios. 
Below we discuss the probability of a companion with the projected separation 
and mass ratio inferred for binary source models. 

The projected separation of binary source components is 
$\sqrt{\left(\frac{t_{0,2}-t_{0,1}}{t_{\rm E}}\right)^2+u_{0,1}^2} = 4.4$ 
times $\hat{r}_{\rm E}$ (Einstein ring radius projected on the source plane).  
The $\hat{r}_{\rm E}$ is 
$2.7_{-1.2}^{+1.6}~{\rm AU}$ for the double-source model, 
leading to a very wide distribution of binary source projected separations.  
The first source is close to the RC on the CMD, and we may guess that 
the first-source mass is similar to that of typical RC stars, which is on 
the order of $0.9~{\rm M_{\odot}}$.  The mass of the RC stars was estimated 
based on \citet{nataf12a} and the mass-loss
rates from \citet{miglio12}.  The isochrones suggest the secondary source 
mass of $0.35~{\rm M_{\odot}}$, hence, the total mass of the binary source is 
$1.25~{\rm M_{\odot}}$ and the mass ratio is $0.4$.  
The mean value of $\hat{r}_{\rm E}$ multiplied by $4.4$, corrected 
for a projection factor of $\sqrt{3/2}$, and combined with the total mass of 
the system results in an orbital period of $52~{\rm years}$ or 
$\log{P~{\rm [day]}} = 4.3$.  This is close to the peak of period distribution 
of local binary systems derived by \citet{raghavan10} of $5.0\pm2.3$.  
Thus, the inferred properties of the binary source are not, 
considered in isolation, particularly unusual. 

Nevertheless, our analysis suggests that a binary lens model is more probable than 
the binary source model, 
because the source can pass farther away from the planetary caustic to 
produce the observed signal than the lens can pass from the second source.  
In other words, we find that the prior phase space to produce the secondary 
signal seen is considerably larger for the single source binary lens model than for 
the binary source single lens model.  
To quantify this argument, we selected all the wide binary lens and 
binary source models with $\Delta\chi^2 < 9$ relative to the best-fitting 
model in each group.  For each model we calculated the minimum approach 
distance to the planetary 
caustic or the second source and divided it by the distance from 
the primary lens/source.  The range of derived values is larger by 
a factor of $3.9$ for the binary lens model than for the binary source model, 
indicating that the source can pass considerably farther away from the planetary 
caustic than the lens can pass from the secondary source to produce the observed signal.

We combine the binary source orbital period and assume a circular orbit
to derive the brighter component radial velocity 
semi-amplitude of $K\sim 2.3\sin i~{\rm km~s^{-1}}$, which can be detected 
by the long-term spectroscopic follow-up.  However, the range of orbital periods, 
and hence radial velocity semi-amplitude, is wide and inclination is unknown.   
The lack of detection of 
change in the radial velocity would not necessarily be conclusive in deciding whether 
the binary lens or the binary source model is true. 

Assuming heretofore that the wide binary lens model is the correct one, there 
is only one microlensing planet with a larger inferred value of $s$: 
OGLE-2008-BLG-092LAb.  The binary lens versus binary source degeneracy was 
resolved for OGLE-2008-BLG-092 using the light-curve analysis alone 
but was based on two important factors: the
presence of a third microlensing body in that system, and the fact that the planetary subevent 
amplitude was larger than the host subevent amplitude 
\citep{poleski14c}.  

Two other events suffer from a severe degeneracy caused 
by poor coverage of the anomaly observed far from the peak of the event: 
MACHO-97-BLG-41 \citep{bennett99,albrow00,jung13} and 
OGLE-2013-BLG-0723 \citep{udalski15c,han16c}.  In both these cases 
the degeneracy was between the binary lens with an orbital motion and 
the static triple lens models. 

The lack of multiband photometry during the anomaly prevents us from 
conclusively distinguishing between the binary lens and binary source model based 
on the chromaticity of the light curve \citep{gaudi98}.  For binary lens models 
we can predict brightness at the peak 
of the anomaly directly 
and obtain $(V-I)$ of $2.342\pm0.002~{\rm mag}$ for the wide model and 
$2.353\pm0.023~{\rm mag}$ for the close model.  
For the binary source we derived the second-source brightness from the above analysis of 
the source properties and found observed $(V-I)$ color of $3.457$ for 
fiducial properties of the second source.  This allows us to predict 
$(V-I)$ color at the peak of the anomaly of $2.195\pm0.043~{\rm mag}$.  
The color estimates do not use Galactic model priors but take full account 
of the blending flux.  We expect the $(V-I)$ color difference of 
$0.147~{\rm mag}$, which could have been measured if $V$-band photometry had been obtained 
during the anomaly.

% #############################################################################
\section{Ice giant occurrence rates}

\begin{figure}[t!]
%\plotone{plot_s_q_v002_colors.eps}
\plotone{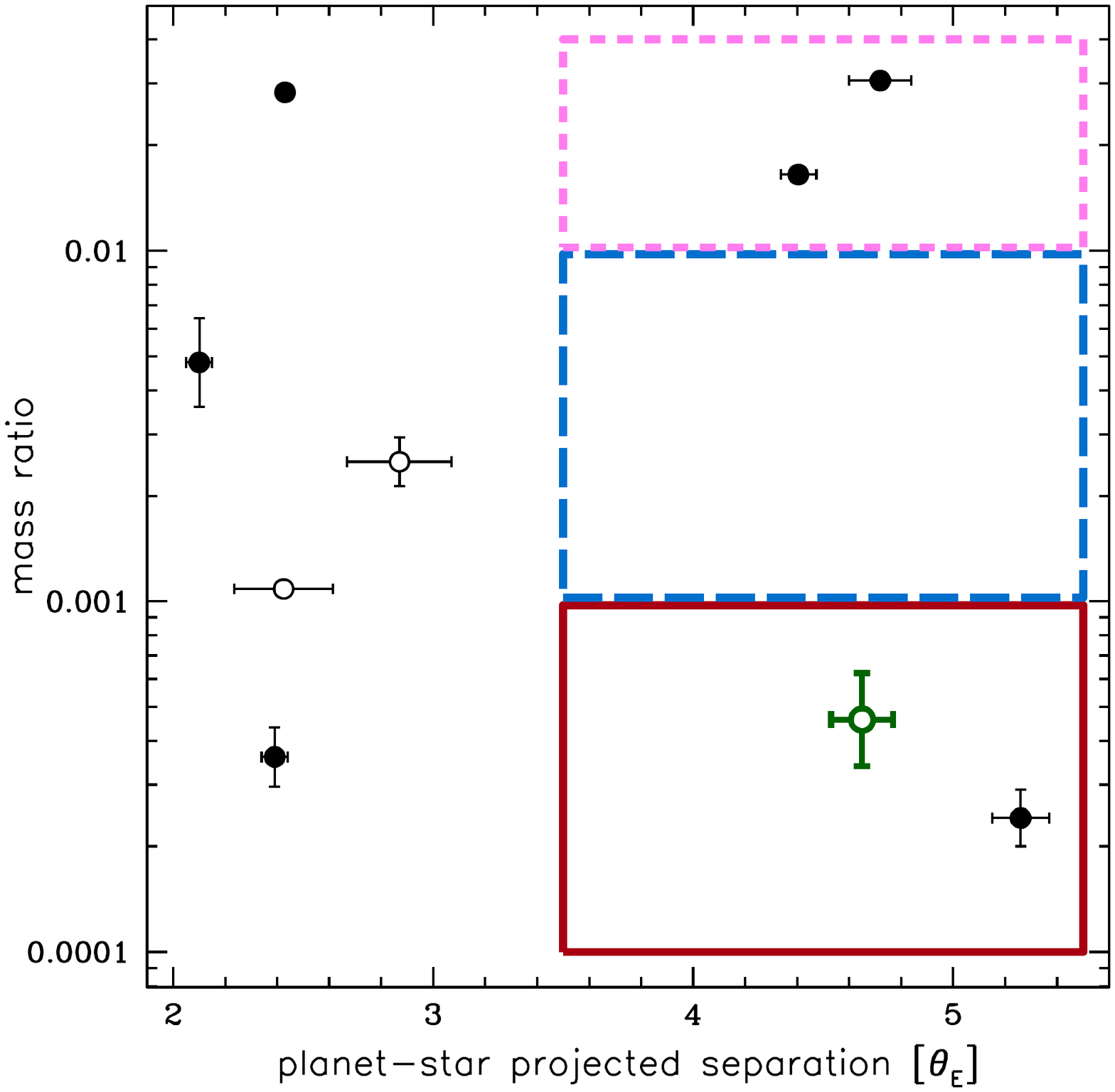}
\caption{Mass ratio vs. projected separation plot for microlensing 
detections with $s>2$.  Open symbols represent objects that suffer 
wide--close (or $s$ vs. $s^{-1}$) degeneracy, while full symbols represent 
objects that are not affected by the wide--close degeneracy. 
Green thick point represents OGLE-2011-BLG-0173.  
The three rectangles show limits used for integrating detection efficiencies 
-- see Table~\ref{tab:sq}.  
The events shown are from the highest to the lowest mass ratio: 
OGLE-2016-BLG-0263 \citep{han17e}, 
OGLE-2014-BLG-1112 \citep{han17d},
MOA-2012-BLG-006 \citep{poleski17}, 
MOA-bin-1 \citep{bennett12}, 
MOA-2007-BLG-400 \citep{dong09b}, 
OGLE-2012-BLG-0563 \citep{fukui15},
OGLE-2011-BLG-0173 (this paper),
MOA-2013-BLG-605 \citep{sumi16b}, and 
OGLE-2008-BLG-092 \citep{poleski14c}.
\label{fig:sq}}
\end{figure}

Here we present an attempt to investigate statistical properties of 
microlensing planets on wide orbits using currently available data.  
In Figure~\ref{fig:sq} we show mass ratio versus projected separation 
(i.e., the quantities that are measured directly) 
for all low-mass companions with $s>2$.  We see that at the very wide orbits, 
the companions seem to separate into groups: two objects have $q>0.016$,
two others have $q<1.5\times10^{-4}$ and there are no objects detected with mass ratios in between.  
In order to quantitatively assess the number of objects with different 
properties, we divided the detected low-mass companions with separations 
in the $3.5 < s < 5.5$ range into three bins: 
$10^{-4} < q < 10^{-3}$, 
$10^{-3} < q < 10^{-2}$, and 
$10^{-2} < q < 4\times10^{-2}$.  
These bins are marked in Figure~\ref{fig:sq} by thick red, long-dashed blue, 
and short-dashed pink rectangles, respectively.  
We integrated the sensitivity of two recent statistical analyses of 
survey data: \citet{tsapras16} for OGLE-III and \citet{suzuki16} for MOA-II, 
which searched 2433 and 1474 events, respectively.  
Integration was done in each bin separately, assuming one companion 
for each star with given properties.  
\citet{tsapras16} showed their results in physical projected 
separation and companion mass, which we translated back to $(s, q)$ 
assuming $r_{\rm E}=2.3~{\rm AU}$ and $M_l=0.4~{\rm M_{\odot}}$. 
In Table~\ref{tab:sq} we present how many objects each of these surveys 
would detect under our assumptions and how many were in fact detected.  
Additionally, we give the number of objects currently known in each 
bin from all microlensing detections.  The sensitivity of all 
microlensing detections is not known, but it is obviously decreasing with 
decreasing mass ratio.  

There are just a few detections of objects in the considered separation range, 
which prevent us from drawing definitive conclusions.  We note 
that there are many events that were searched for distant companions, which
resulted in only four detections altogether: 
on order of $20,000$ events alerted by the OGLE and MOA surveys, in addition to 
the \citet{tsapras16} and \citet{suzuki16} samples.   
The numbers presented in Table~\ref{tab:sq} suggest that 
companions in the lowest mass ratio range are common \citep[see also][]{foremanmackey16}.  
We can consider them 
as analogs of Uranus and Neptune.  In the highest mass ratio bin, the frequency 
of objects seems significantly smaller.  These objects likely did not form in 
a protoplanetary disk around the primary star and hence probably should be 
considered brown dwarfs 
-- see the discussion in \citet{poleski17} and \citet{bryan17} for more details. 
Note that OGLE-2014-BLG-1112 
with $q=0.028$ and $s=2.4$ has a direct mass measurement of $0.03~{\rm M_{\odot}}$, 
i.e., in the brown dwarf range.  

The intermediate mass ratio range 
$10^{-3}<q<10^{-2}$ has no detections, and it is possible that 
the occurrence rate of these objects is similar to the occurrence
$10^{-2}<q<4\times10^{-2}$ range.  If future data show that the occurrence rate 
in the intermediate mass ratio bin is indeed lower, then in the lowest mass ratio 
bin, than this would indicate the existence of a massive 
ice giant desert.  The objects with $10^{-3}<q<10^{-2}$ have masses of a few 
${\rm M_{Jup}}$, but are on much wider orbits.  For in-situ formation, the lack of such planets 
is naturally understood in the core-accretion planet formation model  
\citep[e.g.,][]{pollack96}.  Protoplanetary disks are likely simply not massive 
enough and the dynamical times are too long at such wide orbits to form Jupiter- 
or higher-mass planets.  

\begin{deluxetable}{l|rr|rr|rr}
\tablecaption{Occurrence rates\label{tab:sq}}
\tablehead{
\colhead{sample} & 
\twocolhead{$10^{-4} < q < 10^{-3}$} & 
\twocolhead{$10^{-3} < q < 10^{-2}$} & 
\twocolhead{$10^{-2} < q < 4\times10^{-2}$} \\
\colhead{} & 
\colhead{detected} & \colhead{expected} & 
\colhead{detected} & \colhead{expected} &
\colhead{detected} & \colhead{expected} 
}
\colnumbers
\startdata
\citet{tsapras16} & $1$ & $4.1$ & $0$ & $28.$ & $0$ & $91.$ \\
\citet{suzuki16}  & $0$ & $1.0$ & $0$ & $7.9$ & $1$ & $52.$ \\
all detections    & $2$\tablenotemark{a} & \nodata & $0$ & \nodata & $2$ & \nodata \\
\enddata
%\tablecomments{Notes.}
\tablenotetext{a}{Here we assume the wide model is correct for OGLE-2011-BLG-0173.}
\end{deluxetable}

% #############################################################################
\section{Summary}

We have presented the analysis of microlensing event OGLE-2011-BLG-0173.  
The event is clearly anomalous, but the physical cause of the anomaly 
is not uniquely determined.  There are two distinct sets of models.  The first 
is a single source and a binary lens, which itself has two families of solutions: 
a less massive companion with a larger projected separation, or a more massive 
companion with a smaller projected separation.  We demonstrated, using the \citet{suzuki16} mass ratio and separation function as a prior, that 
the wide-orbit planet model is more probable.  
Unfortunately, we cannot fully distinguish between the binary lens and binary source 
models, though our considerations support the binary lens model.  
Each of these scenarios demonstrated degeneracies that were uncovered 
for the first time.  We found that the best way for resolving all of these 
degeneracies in future events is to collect continuous,  high-cadence 
photometry, preferably with relatively frequent observations in a second band 
(at least once per day when $q>10^{-4}$ planets are considered).  
This is essentially the strategy that will be employed by the WFIRST microlensing survey. % (Penny et al., in prep).  
The KMTNet survey employs a similar strategy \citep[e.g., ][]{kim18a}, but 
its observations are affected by weather.

The detailed Bayesian analysis provided a resolution of the
model-fitting degeneracies.  In our opinion, more advanced methods, such as 
the hierarchical Bayesian analysis, should be more commonly used in analyzing microlensing events.

We have also analyzed the set of known low-mass microlensing companions with 
large projected separations.  If 
OGLE-2011-BLG-0173 is in fact a binary lens event with a wide-orbit planet,  
then it is only the second low-mass planet on a very wide orbit detected.  
Current surveys have low detection efficiency for wide-orbit planets and 
resulted in two planet detections, suggesting that the wide-orbit planets are not rare.  
There are two other events with wide-orbit companions but their mass ratios are 
larger by $1.5~{\rm dex}$, suggesting that they, perhaps, are more likely binary, 
rather than planetary companions.  The observed lack of detections in 
intermediate mass ratios may be interpreted as  tentative evidence for 
a massive ice giant desert.

% #############################################################################
\acknowledgments
Authors would like to thank Prof.~A.~Gould for consultation,  
C.~Clanton for support in preparing Galactic models, and 
S.~Johnson for comments on manuscript.  
We thank the anonymous referee, whose comments helped to clarify the text. 
OGLE Team acknowledges Profs. M.~Kubiak and G.~Pietrzy\'nski, former members of the team,
for their contribution to the collection of the OGLE photometric data over the past years.
This work was partially supported by NASA contract NNG16PJ32C. 
The OGLE project has received funding from the National Science Centre, Poland,
grant MAESTRO 2014/14/A/ST9/00121 to A.U. 
This research made use of {\sc Astropy}, a community-developed core 
Python package for Astronomy \citep{astropy13}.  

\software{
MultiNest \citep{feroz08,feroz09}, 
%SM \citep{https://www.astro.princeton.edu/~rhl/sm/}, 
SM %\citep{https://www.noao.edu/ctio/sys/manuals/sm/sm.html}, 
Numerical Recipes \citep{press92fortran}}

\bibliography{paper}

\end{document}